\documentclass[11pt]{article}
\usepackage{vietnam}

\def\Journal#1#2#3#4{{#1} {\bf #2}, #3 (#4)}

\def\PRL{\em Phys. Rev. Lett.}
\def\PRD{{\em Phys. Rev.} D}

\def\be{\begin{equation}}
\def\ee{\end{equation}}
\def\bea{\begin{eqnarray}}
\def\eea{\end{eqnarray}}

\newcommand{\Photo}{\includegraphics[height=35mm]{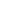}}

\begin{document}
\vspace*{4cm}
\title{THE MOND PHENOMENOLOGY}

\author{B. FAMAEY$^1$ \& S. McGAUGH$^2$}
\address{$^1$Observatoire Astronomique, Universit\'e de Strasbourg, CNRS, UMR 7550, France \\
$^2$Department of Astronomy, Case Western Reserve University, Cleveland, USA}

\maketitle

\abstracts{The $\Lambda$CDM cosmological model is succesful at reproducing various independent sets of observations concerning the large-scale Universe. This model is however currently, and actually in principle, unable to predict the gravitational field of a galaxy from it observed baryons alone. Indeed the gravitational field should depend on the relative contribution of the particle dark matter distribution to the baryonic one, itself depending on the individual assembly history and environment of the galaxy, including a lot of complex feedback mechanisms. However, for the last thirty years, Milgrom's formula, at the heart of the MOND paradigm, has been consistently succesful at predicting rotation curves from baryons alone, and has been resilient to all sorts of observational tests on galaxy scales. We show that the few individual galaxy rotation curves that have been claimed to be highly problematic for the predictions of Milgrom's formula, such as Holmberg~II or NGC~3109, are actually false alarms. We argue that the fact that it is actually possible to predict the gravitational field of galaxies from baryons alone presents a challenge to the current $\Lambda$CDM model, and may indicate a breakdown of our understanding of gravitation and dynamics, and/or that the actual lagrangian of the dark sector is very different and richer than currently assumed. On the other hand, it is obvious that any alternative must also {\it in fine} reproduce the successes of the $\Lambda$CDM model on large scales, where this model is so well-tested that it presents by itself a challenge to any such alternative.}

\section{Introduction}

Assuming our understanding of gravitation and dynamics to be valid on all scales, data ranging from the largest scales to the scales of individual galaxies indicate that the Universe is dominated by a dark sector, composed of dark mater and dark energy. In the current $\Lambda$CDM standard model of cosmology, dark energy is represented by a cosmological constant $\Lambda$ in Einstein's field equations, while dark matter~\cite{Peebles} is a collisionless, dissipationless dust fluid of yet-to-be-discovered stable elementary particles interacting with each other and with baryons (almost) entirely through gravity, immune to hydrodynamical influences (these affect baryons, which can then influence dark matter only through gravitational feedback), and completely unrelated to dark energy.

With these basic ingredients and only six free parameters, this model simultaneously fits no less than 2500 multipoles in the Cosmic Microwave Background (CMB) angular power spectrum, as well as the Hubble diagram of Type~Ia supernovae, the matter power spectrum, and the scale of baryonic acoustic oscillations. Nevertheless these successes rely on a two-components dark sector which awaits independent empirical probes, such as direct detection of particle dark matter in the lab, and the model is unable to predict the very tight scaling relations observed between the distribution of baryons and the gravitational field in galaxies. In fact, a {\it one-to-one} relation between these is observed, independently of the history of individual galaxies. As a matter of principle, this observational fact is difficult to understand in any model based on particle dark matter because the relative distribution of baryons and dark matter is supposed to then depend on the individual assembly history and environment of the galaxy. The observed one-to-one relation corresponds to the empirical formula of Milgrom~\cite{Mil83} who has made a number of {\it a priori} predictions that were verified over the last 30 years \cite{FamMcGaugh}. Most of these succesful predictions involve the ubiquitous appearance of an acceleration constant $a_0 \sim \sqrt{\Lambda}$, whose origin is a deep mystery in the standard context. The success of this formula means that the gravitational field mimicks, for whatever reason, a single effective force law on galaxy scales. This succesful effective dynamics is known as MOdified Newtonian Dynamics or MilgrOmiaN Dynamics (MOND). 

It is important to realize that the {\it a priori} predictions made using the MOND formula are not possible to make {\it at all} in the standard context. Within $\Lambda$CDM, one can only notice the success of the recipe and assume that it is an emergent phenomenon, in the sense of an unequivocal relation emerging from complex phenomena in the baryon physics that one cannot model yet. But to make such a fine-tuning emerge from inherently chaotic and haphazard behaviors, including feedback and very different assembly histories, is far from trivial. So until one has demonstrated how this fine-tuning could occur in the standard context, it is only sensible to consider that this empirically successful recipe might point towards a different theory of the dark sector. 

The latter point of view is the basis of MOND, i.e. hypothesizing that Milgrom's law is built in the fundamental lagrangian of Nature, and that this does not necessarily preclude the dissipationless dust fluid behaviour that one usually attributes to particle dark matter on large scales. The main motivation for building such a theory is thus an {\it empiricist} one, as it is driven by the observed phenomenology on galaxy scales, and not by an aesthetic wish of getting rid of particle dark matter. The real debate is whether Milgrom's law is {\it emergent} or {\it fundamental}.

Hereafter, we first summarize the case for the observational successes of Milgrom's law (Sect.~2), then recall cases  that have been claimed to be highly problematic for MOND and turned out to be false alarms (Sect.~3). We summarize what actually constitute the biggest challenges to the fundamental (rather than emergent) interpretation of the success of this formula (Sect.~4), before concluding (Sect.~5).

\section{The case for MOND}

The simplest formulation of the MOND paradigm is that the total gravity ${\bf g}$ within a galaxy is related to the baryon-generated one ${\bf g_N}$ by Milgrom's formula:
\begin{equation}
{\bf g} = \nu\left(\frac{g_N}{a_0}\right){\bf g_N}, 
\label{inversemoti}
\end{equation}
where 
\begin{equation}
\nu(x) \rightarrow 1 \; {\rm for} \; x \gg 1 \; {\rm and} \; \nu(x) \rightarrow x^{-1/2} \; {\rm for} \; x \ll 1.
\end{equation}
A function which is known to work very well observationally is, for instance~\cite{FamMcGaugh}:
\begin{equation}
\nu(x) = (1-e^{-x^{1/2}})^{-1}.
\end{equation}
In order for ${\bf g}$ to remain a conservative force field, the above expression cannot be exact outside of highly symmetric situations, but it allows us to make very general predictions for galactic systems, i.e. to derive ``Kepler-like laws'' of galactic dynamics unified by Milgrom's law, just like Kepler laws are unified by Newton's law. These laws of galactic dynamics can be found in Sect. 5.2 of Famaey \& McGaugh~\cite{FamMcGaugh} or in the recent overview by Milgrom~\cite{Millaws}. For instance, these laws imply that the constant $a_0$ defines the transition of the acceleration at which the mass discrepancy between baryonic and dynamical mass appears in the standard framework, defines a critical  surface density for disk stability, and defines the zero-point of the Tully-Fisher relation, a relation which is predicted to be extremely tight, with effectively no scatter, as observed. The observational successes of Milgrom's formula, which implies a one-to-one relation between the distribution of baryons and the gravitational field, are discussed in great details in Famaey \& McGaugh~\cite{FamMcGaugh}. It cannot be stressed enough that Milgrom's formula made most of its successful predictions {\it a priori}, before the actual measurements were carried out. 

To illustrate the success of this formula, let us stress that, for instance, it predicts the {\it shape} of rotation curves as a function of baryonic surface density: High Surface Brightness (HSB) galaxies are predicted to have rotation curves that rise steeply then become flat, or even fall somewhat to the not-yet-reached asymptotic flat velocity, while Low Surface Brightness (LSB) galaxies are predicted to have rotation curves that rise slowly to the asymptotic flat velocity. This is precisely what is observed (see Fig.~1), and this behavior had been predicted by Milgrom before LSB galaxies were even known to exist. Detailed fits to LSB galaxy rotation curves in MOND, using Eq.~3, are displayed on Fig.~2 as an example~\cite{dBM98,FamMcGaugh}.

\begin{figure}
\begin{minipage}{0.9\linewidth}
\centerline{\includegraphics[width=0.9\linewidth]{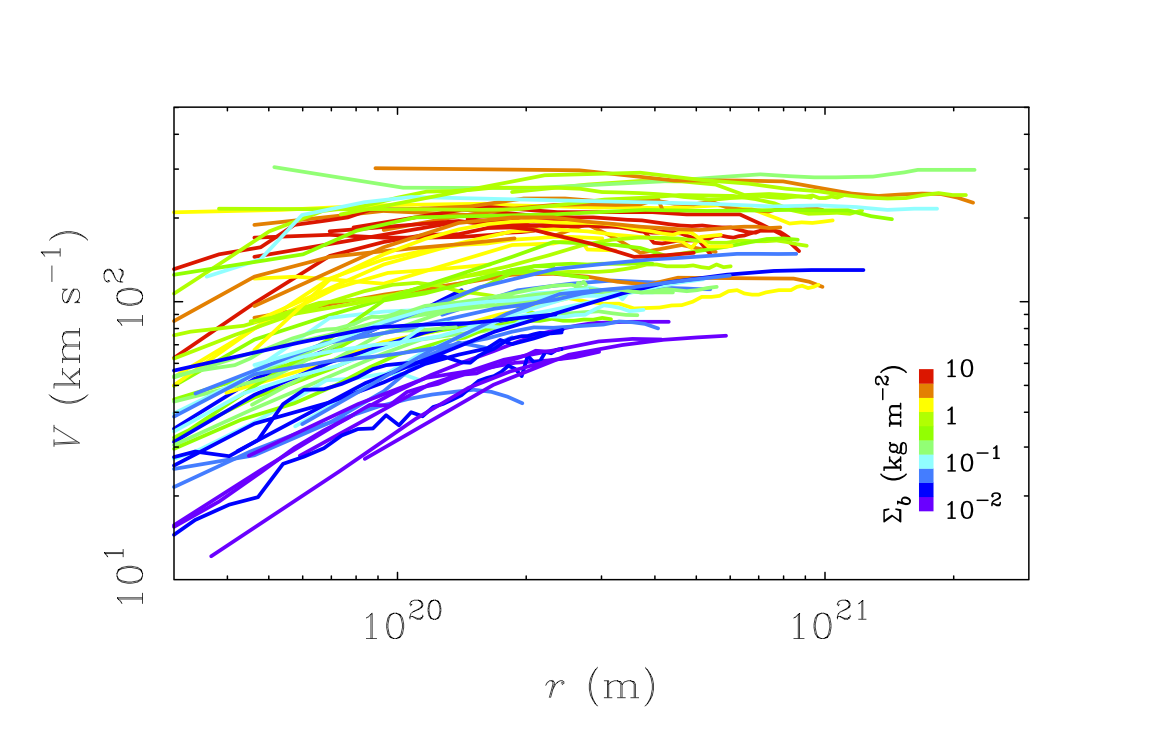}}
\end{minipage}
\caption{The shape of {\it observed} rotation curves (velocity in km/s vs. radius in m) depends on surface density exactly as {\it predicted} by MOND. The curves for individual galaxies (solid lines) are color-coded by their characteristic baryonic surface density. In order to be fully independent of any assumption (such as maximum disk) the stellar masses have been estimated here with population synthesis models. High surface density galaxies have rotation curves rising steeply to then become flat, or fall somewhat to their asymptotic flat velocity. Low surface density galaxies have rotation curves rising slowly to the asymptotic velocity. This trend confirms one of the \textit{a priori} predictions of Milgrom's formula and is difficult to understand in $\Lambda$CDM.}
\label{fig:1}
\end{figure}

\begin{figure}
\begin{minipage}{0.9\linewidth}
\centerline{\includegraphics[width=0.9\linewidth]{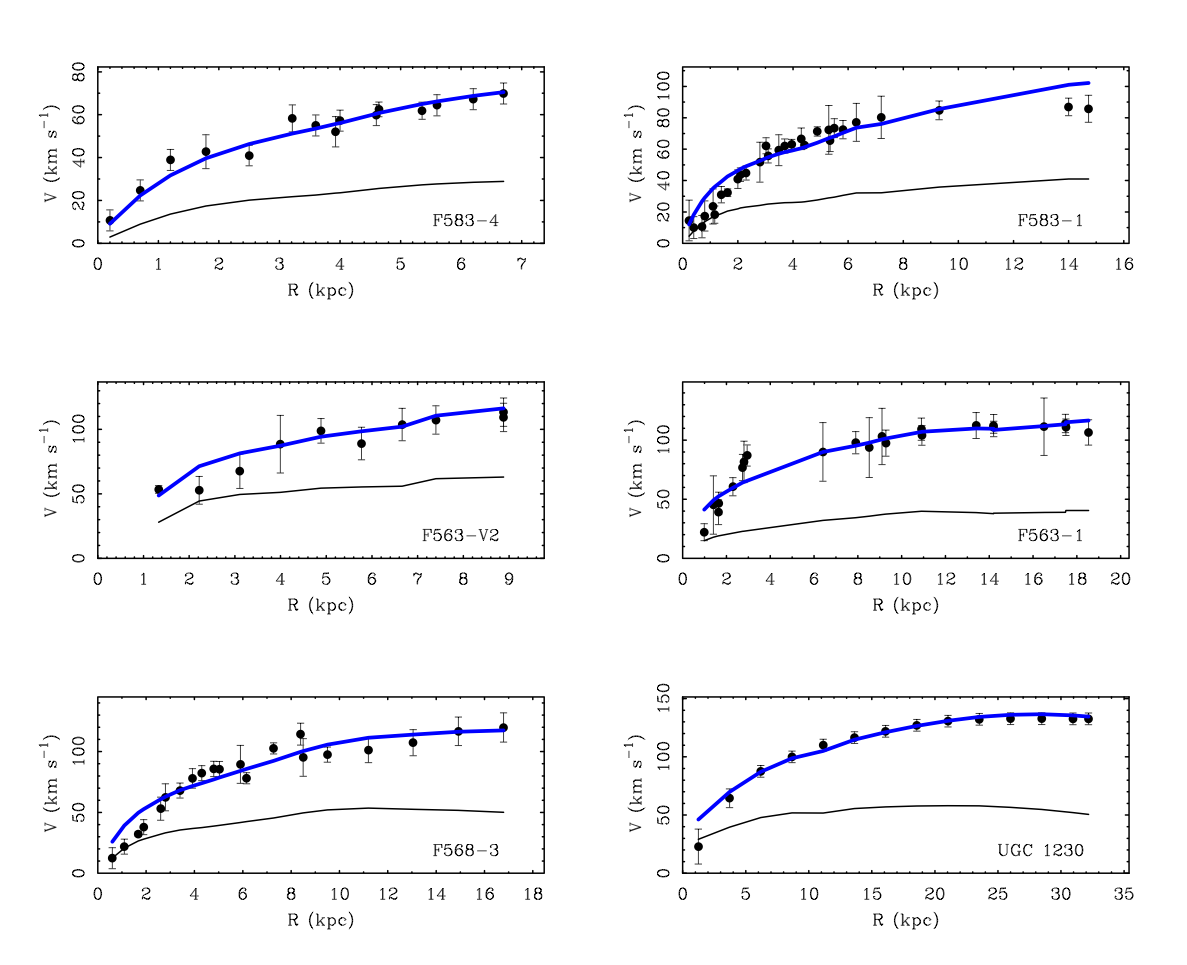}}
\end{minipage}
\caption{MOND rotation curves of LSB galaxies using Eq.~3. All these galaxies have very small accelerations, so that the exact form of the interpolating function does not affect the fits.
}
\label{fig:2}
\end{figure}

Another crucial prediction of MOND is that the mass discrepancy appears whenever the acceleration falls below $a_0$ {\it independently} of the formation scenario of the object. An object born with a high surface density that later becomes more fluffy will in effect see dark matter {\it appear}. Young tidal dwarf galaxies (TDGs) are in this respect an interesting test case. Such galaxies are supposed to form in the collision of larger galaxies. In the standard context, these collisions should be very effective at segregating dark matter particles and baryonic matter: the difference in phase-space distribution of the baryonic disk and dark halo leads to tidal tails that themselves contain very little particle dark matter. When TDGs form from tidal debris, they should thus be largely devoid of such particles. Nevertheless, TDGs do appear to contain enough dark matter to precisely obey the prediction of Milgrom's formula~\cite{GentileTDG}.

A related problem is that the dwarf satellites of the Milky Way and Andromeda are not isotropically distributed around their hosts~\cite{Kroupa,Ibata}. This could indicate that many such dwarf galaxies are not primordial but have been formed tidally in galactic encounters~\cite{Kroupa}. These populations of TDG satellites, associated with globular clusters that formed along with them, would then naturally appear in closely related planes, but to explain why such TDG satellites appear dark matter-dominated, MONDian dynamics would again be the only possible explanation. Interestingly, it was recently shown that, in the MOND context, the currently measured proper motions and radial velocity of M31 would imply that the Milky Way and Andromeda had a past fly-by encounter 7 to 11 Gyr ago before coming to their present positions~\cite{Zhao}. Such an encounter could be the source of many of the dwarf satellites of the two systems.

That dwarf spheroidal galaxies exhibit large mass discrepancies is now well established, but was not anticipated before the fact except by MOND.  Quantitatively, the newly discovered dwarf spheroidal satellites of Andromeda provide another recent example of successful {\it a priori} predictions by MOND. The internal bulk velocity dispersions of individual dwarf satellites were predicted when only photometric data were available~\cite{McMil1}: these predictions turned out to be in accordance with subsequent observations~\cite{Tollerud,Collins,McMil2}.  Among these systems are pairs of similar dwarfs in which one object is effectively isolated while the other is subject to the external field effect imposed by the host galaxy.  MOND makes distinct predictions in these two regimes which the data appear to recognize~\cite{McMil2}.  This distinction should not occur conventionally as the dark matter shields the dwarfs from tidal effects as weak as those predicted to matter by MOND. Making more detailed fits based on the full velocity dispersion profiles of dwarfs can however be difficult due to various uncertainties, notably on velocity anisotropy, so that a large parameter space must be investigated before reaching solid conclusions.

Given all these predictive successes of Milgrom's formula on galaxy scales, one might suspect that it is actually  a fundamental law built in the lagrangian of Nature, or in the fundamental nature of the dark sector. However, as stated above, Milgrom's formula cannot be exact outside of highly symmetric situations, but what is more, it is not relativistic. For 20 years, the main argument against the MOND paradigm has been that there was no relativistic formulation for it, and even that it was maybe not possible to devise one at all. However, during the last 10 years, a plethora of relativistic theories boiling down to Milgrom's formula in the weak-field limit have been proposed~\cite{Bekenstein,Zlosnik,Blanchet,Bimond}. A feature common to all these theories is the addition of new degrees of freedom (scalar fields, vector fields, or a new type of matter such as twin matter or dipolar dark matter) which can play the role of the dust fluid played by plain simple particle dark matter on the  largest scales in $\Lambda$CDM.  It should however be highlighted that these theories cannot be confused with the MOND paradigm itself, which is not a theory {\it per se}. None of these theories can thus be considered to be the final word on the topic.

\section{False alarms for MOND}

There have been, over the last 30 years, a few claims of observations strictly ruling out the predictions of Milgrom's formula in galaxies, which, interestingly, have all turned out to be false alarms. There probably is a lesson to be learned there. Indeed, even the most perfect theory will necessarily show mild discrepancies with some subset of a given large astronomical dataset, and even large discrepancies with a smaller fraction. This is because the data always involve possible systematics or mistaken assumptions. For example, if some distance measurement method has a one-sigma measurement error of 10\%, in one in a hundred cases one will have a much larger error. If the observer just looks at this single case, he might not believe that the error is so large because the method is much more accurate, and therefore he would exclude the theory, while actually all he would have done is pick an outlier which, statistically, should be present within the dataset.

In the case of pressure-supported systems (elliptical galaxies, dwarf spheroidals, globular clusters), the problem  in drawing rigorous conclusions often lies in the uncertainty in the dynamics of stellar tracers, most notably their velocity anisotropy, as well as on the orbit of the object around its host, leading to non-trivial predictions linked to the external field effect. It is thus highly interesting that, for isolated elliptical galaxies, data not plagued by such uncertainties and probing dynamics in the extreme weak-field limit are giving the best possible agreement with MOND: this includes the gravitational field of two ellipticals probed with X-ray emitting gas~\cite{Milellipt} and weak-lensing from a large galaxy sample \cite{Millensing}.

In the case of rotationally-supported systems, such as spiral galaxy rotation curves, the test is much more fruitful as there is no such uncertainty on velocity anisotropy. Observational problems related to the distance and the inclination of the probed galaxy can however remain, as well as extreme sensitivity to the parametrized bulge-disk decomposition, or the presence of non-axisymmetric motions due to, e.g., a central bar. Other problems can also arise, related to the misidentifications of tidal debris for disks in equilibrium, or simply to a misunderstanding of the actual predictions of MOND. We give hereafter a non-exhaustive list of such cases that were deemed problematic for MOND but turned out to be false alarms.

A first example of a misunderstanding of the MOND prediction was the analysis of the HI kinematics of irregular dwarfs~\cite{Lo}, where the MOND baryonic Tully-Fisher relation for the asymptotic circular velocity $v_\infty$ was applied to the half-width at half-maximum of the integrated HI line profile, $v_h$. The mass deduced was systematically smaller than even the HI mass observed in theses systems. However, Milgrom~\cite{MilLo} subsequently showed that applying the correct MOND mass estimator to $v_h$ brought these systems in full agreement with MOND. 

A prototypical example of a problem of distance is the case of the Andromeda IV: when it was thought to be a satellite of M31, the discrepancy with MOND was huge, since it is gas-dominated and has a reasonable inclination. Nevertheless, it was then shown~\cite{Ferguson} that its distance was larger than 5~Mpc, i.e. well outside the confines of the Local Group, bringing it back in accordance with MOND. Actually, the accordance with MOND is now remarkably good; this galaxy has a good inclination and suffers not from uncertainty in the mass-to-light ratio because it is completely gas-dominated~\cite{MilKK}.

An example of misidentification is the HI cloud VIRGOHI21 which had been claimed to constitute a ``dark galaxy" which was the prototypical galaxy-scale counter-example to MOND~\cite{Funkhouser}. This cloud was however subsequently shown \cite{BDucfake} to be a tidal debri from a high velocity collision, and not a rotationally supported galaxy, thus posing no challenge to MOND.

NGC~7814 is on the other hand a good example of the sensitivity to the bulge-disk decomposition. Having at first been claimed to be highly problematic because it needed an unrealistically high mass-to-light ratio for the bulge~\cite{Filippo}, it was subsequently shown to be in good agreement with MOND by virtue of the use of a double-S\'ersic fit to the bulge profile~\cite{angus7814}.

A good example of imprecise inclination causing trouble is the dwarf galaxy Holmberg~II within the M81 galaxy group. At first claimed to rule out MOND with high significance~\cite{Sanchez1,Sanchez2}, it was subsequently shown, by remodelling its HI data cube, that its inclination was closer to face-on than previously derived~\cite{GentileHolmberg}. This implies that Holmberg II has a higher rotation velocity in its outer parts than previously assumed, which, although not very precisely constrained, is fully consistent with the MOND prediction.

The most recent example of a galaxy claimed to be problematic for MOND is the Magellanic-type spiral NGC~3109, in the outskirts of the Local Group. This galaxy has been claimed to be problematic for a long time, and it is in this sense interesting that the recently measured more accurate gas distribution~\cite{Carignan} has reduced the originally claimed discrepancy. Actually, a close look at the top row of Fig.~15 of Carignan et al.~\cite{Carignan}, where $a_0$ takes a standard value, reveals an excellent fit to the outer parts of the rotation curve. The two first data points, at $R<3$~kpc however exhibit a too low rotational velocity compared to the MOND prediction. This, however, can be explained by the non-circular motions associated with the weak bar at the center, as shown in Fig.~5 of Valenzuela et al.~\cite{Valenzuela}, meaning that NGC~3109 is certainly not problematic for MOND.

In conclusion, all the cases of rotationally-supported systems that have been claimed to be highly problematic for MOND have turned out to be false alarms. There nevertheless remain a few real borderline cases. The most relevant one is probably NGC~3198, for which the quality of the MOND fit is merely modest. It is in this sense interesting to compare the fits obtained with a high-resolution survey (THINGS) and with a survey more sensitive to the outer, fainter emission (HALOGAS)~\cite{Gentile3198}: in the inner parts MOND fits the THINGS rotation curve better than the HALOGAS one, and in the outer parts the situation is reversed, thus meaning that MOND fits best where each rotation curve is expected to give the most reliable result. Small non-axisymmetric motions and uncertainty on the distance might also play a role. Investigating the distance of this galaxy by different methods than the Cepheid one, which might be affected by some systematics, would thus be of the highest interest.

\section{Real challenges for MOND}

While the MOND paradigm is extremely succesful and predictive on galaxy scales, where the $\Lambda$CDM model is currently only able to make rather unsuccesful {\it postdictions}, the situation is actually reversed on the largest scales. The MOND paradigm is {\it a priori} mute on cosmology: only its currently proposed parent relativistic theories are able to make predictions there, and the hope would be that their new degrees of freedom would be able to play the role of the dust fluid played by particle dark matter in $\Lambda$CDM. But focusing on such proposed relativistic MOND theories, it turns out that it is very difficult for many of them to reproduce the CMB angular power spectrum~\cite{Zuntz} without leading to an exaggerated integrated Sachs-Wolfe effect. Some theories \cite{Bimond} have however not been tested yet on CMB scales.

A related but somewhat independent problem is that Milgrom's formula fails in galaxy clusters~\cite{Gerbal,Sanders}. This might simply indicate that more baryons need to be found there, but these baryons should be in collisionless form~\cite{Angusbullet}, such as cool and compact clouds of molecular gas~\cite{MilCBDM}, and one should understand whether such objects could be stable and why they would be found only in galaxy clusters but not in galaxies. Another possibility is to resort to additional non-baryonic particles of hot dark matter, which would not condense on galaxy scales but could play a dominant role in galaxy clusters. Combining such a hot dark matter component with MOND could additionally explain for free the high third peak in the angular power spectrum of the CMB~\cite{AngusCMB}, but at the price of leading to a dramatic overproduction of high-mass galaxy clusters in the process of structure formation~\cite{AngusSim}. Finally, the most satisfying explanation would perhaps be that the new degrees of freedom in relativistic theories play the role of collisionless particle dark matter both on CMB and cluster scales. 

Perhaps the most advanced theory in that respect is that of ``dipolar dark matter"~\cite{Blanchet}: this theory relies  on the existence of a dark fluid endowed with a gravitational dipole moment vector which effectively modifies gravity. This is very different from the usual particle dark matter hypothesis. If the dipolar fluid is at rest with respect to a given galaxy, and if its monopolar density is much smaller than the baryonic density, Milgrom's law is recovered. The $\Lambda^2 \sim a_0$ coincidence is also explained naturally. Thanks to the monopolar density of the fluid, the theory reproduces all successes of $\Lambda$CDM at linear order for the expansion, for large scale structure formation, and for the CMB (naturally explaining the high third peak in the angular power spectrum). Nevertheless, this model is far from perfect: it exhibits an instability which is fundamentally unsatisfying, even though the unstable modes develop on long timescales comparable to a Hubble time. Moreover, as the dipolar fluid does not move on geodesics, and thereby effectively breaks the weak equivalence principle, it is not clear that a dwarf satellite galaxy could remain stable while orbiting around its host. Finally, the residual missing mass in galaxy clusters is concentrated in the central parts of these objects, whilst the dark fluid is supposed to cluster only weakly, so it is not clear that it can solve the missing mass problem in clusters. All these questions should be answered through the advent of large numerical simulations of structure formation within this framework. This is also the case for the other existing MOND theories and for all those that might be developed in the future. One cannot hope to make a fair comparison of these theories with $\Lambda$CDM until a comparable amount of effort has been made to explore their predictions.

\section{Conclusion}

It is clear that we nowadays have solid evidence for something behaving as a dissipationless dust fluid on the largest scales (the high 3rd peak of the CMB being its most obvious manifestation), a role played by particle dark matter in the current $\Lambda$CDM model. However, there is {\it zero} experimental evidence that the answer is a fluid of stable elementary dark particles interacting with each other and with baryons almost entirely through gravity, without any additional fundamental property encoded in the lagrangian. Actually, as we have shown here, one could even argue that there is evidence of the contrary, through all the galaxy-scale regularities that have been predicted 30 years ago by Milgrom and that are impossible to predict in the particle dark matter context. On the other hand, it is obvious that any alternative will also have to reproduce the successes of the $\Lambda$CDM model on large scales, where this model is so well-tested that it presents by itself a challenge to any such alternative. 

In conclusion, whatever one's interpretation of the success of Milgrom’s formula, it is only fair to recognize a rather successful and predictive phenomenological law for galaxies has been uncovered 30 years ago. Of course it is not completely impossible in physics to have some regularity emerging from very complex physical processes such as baryonic feedback, but to make such processes erase completely the effect of different individual assembly histories is {\it far} from trivial. So until one could show how the emergence of such a tight relation would be possible in the particle dark matter context, it is only natural to also look for more fundamental explanations, at least until particle dark matter is actually detected in the laboratory.


\section*{References}

\begin{thebibliography}{99}

\bibitem{Peebles}J. Peebles, \Journal{Essay for the Dark Matter Sackler Colloquium}{}{arXiv:1305.6859}{2013}.

\bibitem{Mil83}M. Milgrom, \Journal{\it Astrophys. J.}{270}{365}{1983}.

\bibitem{FamMcGaugh}B. Famaey and S. McGaugh, \Journal{\it Living Reviews in Relativity}{15}{10}{2012}.

\bibitem{Millaws}M. Milgrom, \Journal{e-print}{}{arXiv:1212.2568}{2012}.

\bibitem{dBM98}W.J.G. de Blok and S. McGaugh, \Journal{\it Astrophys. J.}{508}{132}{1998}.

\bibitem{GentileTDG}G. Gentile, \Journal{\it Astron. Astrophys.}{472}{L25}{2007}.

\bibitem{Kroupa}P. Kroupa {\it et al}, \Journal{\it Astron. Astrophys.}{523}{A32}{2010}.

\bibitem{Ibata}R. Ibata {\it et al}, \Journal{\it Nature}{493}{62}{2013}.

\bibitem{Zhao}H.S. Zhao {\it et al}, \Journal{\it Astron. Astrophys.}{557}{L3}{2013}.

\bibitem{McMil1}S. McGaugh and M. Milgrom, \Journal{\it Astrophys. J.}{766}{22}{2013}.

\bibitem{Tollerud}E. Tollerud {\it et al}, \Journal{\it Astrophys. J.}{768}{50}{2013}.

\bibitem{Collins}M. Collins {\it et al}, \Journal{\it Astrophys. J.}{768}{172}{2013}.

\bibitem{McMil2}S. McGaugh and M. Milgrom, \Journal{\it Astrophys. J.}{775}{139}{2013}.

\bibitem{Bekenstein}J. Bekenstein, \Journal{\PRD}{70}{083509}{2004}. 

\bibitem{Zlosnik}T. Zlosnik {\it et al}, \Journal{\PRD}{75}{044017}{2007}.

\bibitem{Blanchet}L. Blanchet and A. Le Tiec, \Journal{\PRD}{80}{023524}{2009}.

\bibitem{Bimond}M. Milgrom, \Journal{\PRD}{80}{123536}{2009}.

\bibitem{Milellipt}M. Milgrom, \Journal{\PRL}{109}{131101}{2012}.

\bibitem{Millensing}M. Milgrom, \Journal{\PRL}{111}{041105}{2013}.

\bibitem{Lo}K.Y. Lo {\it et al}, \Journal{\it Astron. J.}{106}{507}{1993}.

\bibitem{MilLo}M. Milgrom, \Journal{\it Astrophys. J.}{429}{540}{1994}.

\bibitem{Ferguson}A. Ferguson {\it et al}, \Journal{\it Astron. J.}{120}{821}{2000}.

\bibitem{MilKK}M. Milgrom, , \Journal{e-print}{}{arXiv:1104.1118}{2011}.

\bibitem{Funkhouser}S. Funkhouser, \Journal{\it Mon. Not. Roy. Astron. Soc.}{364}{237}{2005}.

\bibitem{BDucfake}P.-A. Duc and F. Bournaud, \Journal{\it Astrophys. J.}{673}{787}{2008}.

\bibitem{Filippo}F. Fraternali {\it et al}, \Journal{\it Astron. Astrophys.}{531}{A64}{2011}.

\bibitem{angus7814}G. Angus {\it et al.}, \Journal{\it Astron. Astrophys.}{543}{A76}{2012}.

\bibitem{Sanchez1}F.J. S\'anchez-Salcedo and A.M. Hidalgo-G\'amez, \Journal{e-print}{}{arXiv:1105.2612}{2012}.

\bibitem{Sanchez2}F.J. S\'anchez-Salcedo {\it et al}, \Journal{\it Astron. J.}{145}{61}{2013}.

\bibitem{GentileHolmberg}G. Gentile {\it et al}, \Journal{\it Astron. Astrophys.}{543}{A47}{2012}.

\bibitem{Carignan}C. Carignan {\it et al}, \Journal{\it Astron. J.}{146}{48}{2013}.

\bibitem{Valenzuela}O. Valenzuela {\it et al}, \Journal{\it Astrophys. J.}{657}{773}{2007}.

\bibitem{Gentile3198}G. Gentile {\it et al}, \Journal{\it Astron. Astrophys.}{554}{A125}{2013}.

\bibitem{Zuntz}J. Zuntz {\it et al},  \Journal{\PRD}{81}{104015}{2010}.

\bibitem{Gerbal}D. Gerbal {\it et al}, \Journal{\it Astron. Astrophys.}{253}{77}{1992}.

\bibitem{Sanders}R.H. Sanders, \Journal{\it Astrophys. J.}{512}{L23}{1999}.

\bibitem{Angusbullet}G. Angus {\it et al}, \Journal{\it Astrophys. J.}{654}{L13}{2007}.

\bibitem{MilCBDM}M. Milgrom, \Journal{\it New Astron. Review}{51}{906}{2008}.

\bibitem{AngusCMB}G. Angus,  \Journal{\it Mon. Not. Roy. Astron. Soc.}{394}{527}{2009}.

\bibitem{AngusSim}G. Angus {\it et al}, \Journal{\it Mon. Not. Roy. Astron. Soc.}{}{arXiv:1309.6094}{2013}.

\end{thebibliography}
\end{document}